\documentstyle[11pt,a4,%ere99,
amssymb]{article}
\input{amssym.def}
\input{amssym.tex}
\newcommand{\beq}[1]{\begin{equation}\label{#1}}
\newcommand{\eeq}{\end{equation}}
\newcommand{\bear}[1]{\begin{eqnarray}\label{#1}}
\newcommand{\ear}{\end{eqnarray}}
\newcommand{\bearr}[1]{\begin{eqnarray}\label{#1}\lal}

\newcommand{\nn}{\nonumber}

\def\nnnv{\nonumber\\[5pt] \lal }

\newcommand{\rf}[1]{(\ref{#1})}

\def\nhq{\hspace{-0.5em}}
\def\nqq{\hspace{-2em}}
\def\al{&\nhq}
\def\lal{&&\nqq}               % left alignment
\def\eql{\al=\al}

\def\cm{\hspace{1cm}}
\def\inch{\hspace{1in}}

\def\yyy{\\[5pt] \lal}

\def\e{{\rm e}}
\def\m{{\rm m}}
\def\E{{\rm E}}

\def\Qie{Q_{\e I}}
\def\Qim{Q_{\m I}}

\def\half{{\frac{1}{2}}}

\newcommand{\vol}{\mbox{\rm vol}}

\newcommand{\p}{\partial}

\newcommand{\sq}[1]{\sqrt{|#1|}}
\newcommand{\ints}{ \mbox{\rm int} }

\def\bhs{black holes}%
\newcommand\oG{\overline{G}}
\newcommand{\ol}{ \overline}

\newcommand\oI{\overline{I}}
\newcommand\eqdef{\stackrel{\rm def}{=}}

\newcommand{\nl}{ {\hfill \break} }

\renewcommand{\ol}{ \overline}

\newcommand{\inc}{ {\hookrightarrow } }
\newcommand{\bdy}{ {\partial } }

\newcommand{\Ric}{ {\rm Ric} }

\newcommand{\Diff}{ {\rm Diff} }

\newcommand{\GHY}{ {\rm GHY} }

\newcommand{\diag}{ {\rm diag} }

\newcommand{\sign}{ {\rm sign} }
\newcommand{\pr}{ {\rm pr} }
\newcommand{\eps}{ \varepsilon }

\newcommand\mustbe{\stackrel{!}{=}}

\newcommand{\tr}{\mbox{\rm Tr}\,}

\newcommand{\Cinf}{C^{\infty}}

\def\const{{\rm const}}

\def\mO{{[-1]\mathstrut}}
\def\pO{{[1]\mathstrut}}

\newcommand{\T}{{\frak T}}
\newcommand{\N}{ \mbox{\rm I$\!$N} }
\newcommand{\R}{ \mbox{\rm I$\!$R} }
\def\C{\mbox{\rm {I\kern-.520em C}}}
\begin{document}

\centerline{\Large\bf Multidimensional $\sigma$-Model with Black Holes and p-Branes
\footnote{presented at the Spanish Relativity Meeting, ERE 99, Bilbao}
}
\centerline{\large\bf Martin Rainer}

\begin{center}
Mathematische Physik I, Institut f\"ur Mathematik,  Universit\"at Potsdam,
\\
PF 601553, D-14415 Potsdam, Germany
\\
E-mail: {\em {\tt mrainer@rz.uni-potsdam.de}}
\end{center}

\abstract{
The bosonic content of string theory in curved background of
multidimensional structure with $p$-branes has a systematic geometrical
description as an effective $\sigma$-model of gravity in lower dimension (say $3+1$)
with additional interacting dilatonic and $p$-brane fields.
%(see gr-qc/9812031).
%%
%In gr-qc/9608020 the sigma-model description has been obtained for
%a multidimensional curved space-time background.
%Corresponding solutions with $p$-branes
%corresponding to bosonic sectors of perturbative M-theory ($10$-dim.
%string theories and $11$-dim. supergravity) and others have been obtained
%first in gr-qc/9705005.
%
If the  target-space is locally symmetric, solutions with
intersecting $p$-branes can be found.  Some static solutions are $p$-brane
generalizations of black holes (including the standard Reissner-Nordstr\"om class),
which allow the prediction of detectable features of the higher-dimensional
$p$-brane geometry via scaling properties of black hole thermal properties.
E.g. the Hawking temperature $T_H$
depends critically on the $p$-brane intersection topology.
}

%%%%%%%%%%%%%%%%%%%%%%%%%%%%%%%%%%%%%%%%%%%%%%%%%%%%%%%%%%%%%%%%%%%%%%%%%%
\section{\bf Physical motivation}
\setcounter{equation}{0}
%%%%%%%%%%%%%%%%%%%%%%%%%%%%%%%%%%%%%%%%%%%%%%%%%%%%%%%%%%%%%%%%%%%%%%%%%
The purpose of this paper is to clarify the geometric structure of
the effective $\sigma$-model for multidimensional Einstein
geometry described recently.\cite{Ra99}
It is an extension of an early $\sigma$-model
for pure multidimensional geometry\cite{RZ96,RZ98}
which by now also admits solutions with
intersecting electric $p$-branes\cite{IMR} or intersecting
electric and magnetic black branes\cite{BKR} (generalized black holes).

On the physical side, it is sufficiently predictive to admit tests
of the multidimensional geometric models
down to their concrete physical imprints in the physical
space-time and, on the mathematical side,
it provides a rigorous setting based on a well-defined class
of higher-dimensional geometries.
Applications include such different
directions as cosmology, (extended) string theory, and
quantization of certain higher-dimensional geometric actions.
For cosmology it provides a unified view of spatially homogeneous
and inhomogeneous models, and it enables a systematic adjustment
to the physical frame of cosmological predictions (such as inflation)
within a multidimensional setting.\cite{RZ98}
For M-theory and string theory generalizations,
the $\sigma$-model provides a purely geometric understanding
of typical objects such as $p$-branes
without any recursion to SUSY or duality,
and  allows
to relate the $D$-dimensional geometric content
to features of a physical space-time
of dimension $D_0=1+3$.
Last not least, it provides a useful framework on the
(the multidimensional) class of $D$-geometries,
which either can be subject to covariant quantization,
or to canonical quantization (whenever this possible for
its geometrical base in dimension $D_0=4$).
%%%%%%%%%%%%%%%%%%%%%%%%%%%%%%%%%%%%%%%%%%%%%%%%%%%%%%%%%%%%%%%%%%%%%%%%
\section{\bf Multidimensional geometry}
\label{Sect. 1}
\setcounter{equation}{0}
%%%%%%%%%%%%%%%%%%%%%%%%%%%%%%%%%%%%%%%%%%%%%%%%%%%%%%%%%%%%%%%%%%%%%%%%
%For the purpose of this paper let
A ($\Cinf$-) {\em multidimensional} (MD) manifold $N$
be topologically defined by
a $\Cinf$-bundle
\beq{1.1}
M \inc N \rightarrow \overline{M}_{0}
\eeq
with a direct product
\beq{1.x0}
M:=\times_{i=1}^{n} M_{i}
\eeq
of internal $\Cinf$ factor spaces $M_i$, $i=1,\ldots,n$,
as a standard fiber,
and a distinguished  $\Cinf$ base manifold $\overline{M}_{0}$.
The MD manifold $N$ is called {\em internally homogeneous}
if there exists a direct product group
$G:=\bigotimes_{i=0}^{n} G_{i}$
with a direct product realization $\tau:=\otimes_{i=0}^{n} \tau_i$
on $\Diff(M):=\bigotimes_{i=0}^{n}\Diff(M_i)$ such that
for $i=0,\ldots,n$ the realization
\beq{1.x1}
\tau_i:G_{i}\to \Diff(M_i)
\eeq
yields a transitive action of $\tau_i(G_{i})$ on $M_i$.
%
%\eop

For $i=0,\ldots,n$, let each factor space $M_i$ be equipped
with a smooth homogeneous metric $g^{(i)}$.
%Furthermore,
Let
$\overline{M}_{0}$ be equipped with an arbitrary $\Cinf$-metric
$\overline{g}^{(0)}$, and let $\ol\gamma$ and $\beta^i$ ,
$i=1,\ldots,n$ be smooth scalar fields on $\overline{M}_{0}$.
%
%Then,
Under the projection
$\pr: N\to \overline{M}_{0}$
a pullback
of $e^{2\ol\gamma}\overline{g}^{(0)}$
from $x\in \overline{M_{0}}$ to $z\in\pr^{-1}\{x\}\subset M$,
consistent with the fiber bundle \rf{1.1}
and the homogeneity of internal spaces,
is given by
\beq{1.2}
g_{(z)}:=e^{2\ol\gamma(x)}\overline{g}^{(0)}_{(x)}
\oplus_{i=1}^{n} e^{2\beta^i(x)}g^{(i)} .
\eeq
The function $\ol\gamma$
fixes a {\em gauge} for the (Weyl)
{\em conformal frame} on $\overline{M_0}$,
corresponding just to a particular choice
of geometrical variables.
For example $\ol\gamma:=0$ defines the
Brans-Dicke frame.
\rf{1.2} is a multidimensional generalization of
a warped product\cite{BeemEE} $N=\ol{M}_0\times_a M$
where $a:=e^\beta$ is
now a {\em vector-valued} root warping function.
We also define
\beq{e14}
         \eps (I) := \prod_{i\in I} \eps_i  ;       \cm
         \sigma_0 := \sum_{i=0}^{n} d_i \beta_i  ,  \cm
         \sigma_1 := \sum_{i=1}^{n} d_i \beta_i  ,  \cm
        \sigma(I):= \sum_{i\in I} d_i \beta_i  ,
\eeq where $\eps_i:=\sign(|g^{(i)}|)$ and $M_i\subset M$ for
$i=0,\ldots,n$ are all homogeneous factor spaces.
The Levi-Civita connection $\Gamma$ corresponding to \rf{1.2}
does {\em not} decompose multidimensionally, and
neither does the Riemann tensor.
The latter is a section in $\T^1_3M$
which is not given as a pullback to $\ol M_0$ of a section in
the direct sum $\oplus_{i=1}^{n} \T^1_3M_i$
of  corresponding tensor bundles over the factor manifolds.
However, with (\ref{1.2})
the Ricci tensor decomposes again multidimensionally:
\beq{1.2R}
\Ric[g]=
\Ric^{(0)}[g^{(0)}, {\ol\gamma}
%,\p g^{(0)}, \p {\ol\gamma}, \p\p g^{(0)}, \p\p {\ol\gamma}
; \phi%,\p \phi,\p \p \phi
]
\oplus^{n}_{i=1}
\Ric^{(i)}[g^{(0)}, {\ol\gamma}%
%, \p g^{(0)}, \p {\ol\gamma}
; g^{(i)},\phi%,\p g^{(i)},\p \phi,\p\p g^{(i)},\p\p \phi
] .
\eeq

\beq{2.5}
f \equiv {f}[\ol{\gamma}, \beta]
:= ({D}_0 - 2) \ol{\gamma} +\sum_{j=1}^{n} d_j  \beta^j
\eeq
can be resolved for
if and only if ${D}_0 \neq 2$.
%Then,
%\rf{2.4} can also be written as
The Ricci curvature scalar then satisfies
\bear{2.6}
 R[g] \! &\!-\! &\!
e^{-2\ol{\gamma} } R[\ol{g}^{(0)}]
-\sum_{i=1}^{n} e^{-2\beta^i} R_i \ =
\\\nn
=\! &\!-\! &\!  e^{-2\ol{\gamma}}
\left\{
\sum_{i=1}^{n} d_i (\p\beta^i)^2
+ ({D}_0 - 2) (\p \ol{\gamma})^2 - (\p f) \p (f + 2 \ol{\gamma})
+ R_{B} \right\} ,
\\
\label{2.7}
R_B &:=& \frac{1}{\sq {\ol{g}^{(0)}} } e^{-f}
     \p_{\mu} \left[2 e^f \sq {\ol{g}^{(0)} }
     {\ol{g}^{(0)\mu \nu}} \p_{\nu} (f + \ol{\gamma}) \right] ,
\ear
where the last term will yield just a boundary contribution \rf{2.12}
to the action \rf{2.11} below.
Let all $M_{i}$, $i=1,\ldots,n$,  be connected and oriented.
\beq{2.9}
\mu:= \prod_{i=1}^n \mu_i, \quad
\mu_i := \int_{M_i} \tau_i =\int_{M_i} \vol(g^{(i)})  ,
\eeq
the total internal volume resp. an internal factor's volume.
$R$-homogeneity of $g^{(i)}$
(in particular satisfied for Einstein spaces),
guarantuees that the ratios
%\beq{ratios}
$\frac{\rho_i}{\mu_i}=R[g^{(i)}] $,
%\eeq
$i=1,\ldots,n$, are just finite constants.
The
$D$-dimensional coupling constant $\kappa$
can be tuned such that
%,
%under the dimensional reduction $\pr: M\to \ol{M}_0$,
\beq{2.10}
\kappa_0:=\kappa \cdot \mu^{-\frac{1}{2}}
\eeq
becomes the ${D}_0$-dimensional physical coupling constant.
If ${D}_0=4$, then ${\kappa_0}^2=8\pi G_N$.
%%%%%%%%%%%%%%%%%%%%%%%%%%%%%%%%%%%%%%%%%%%%%%%%%%%%%%%%%%%%%%%%%%%%%%%%
\section{\bf The effective $\sigma$-model}
\label{Sect. 2}
\setcounter{equation}{0}
%%%%%%%%%%%%%%%%%%%%%%%%%%%%%%%%%%%%%%%%%%%%%%%%%%%%%%%%%%%%%%%%%%%%%%%%
With the total dimension $D$ and $\kappa^2$ a $D$-dimensional
gravitational constant, we consider a purely gravitational
action of the form
\beq{2.11}
S = \frac{1}{2\kappa^2} \int_{N} d^{D}z \sq g
\{ R[g] %- 2\Lambda
\} + S_{\rm GHY} %+ S_{\Phi}+S_{\rho}
 .
\eeq
Here a (generalized) Gibbons-Hawking-York \cite{GH,Y} type
boundary contribution $S_{\rm GHY}$ to the action is taken
to cancel boundary terms.
Eqs.\rf{2.6} and \rf{2.7} show that $S_{\rm GHY}$ should be taken in the form
\beq{2.12}
S_{\rm GHY} := \frac{1}{2\kappa^{2}} \int_{N} d^{D}z \sq g
     \{ e^{-2\ol \gamma} R_{B} \} .
\eeq
After dimensional reduction the action \rf{2.11} reads
\bear{2.15}
S=\frac{1}{2\kappa^2_0}
\int_{\ol{M}_0}d^{{D}_0}x
\sqrt{|{\ol{g}^{(0)}}|}
%&e^f&
e^f
\left\{
R[\ol{g}^{(0)}]
+(\p f)(\p[f+2\ol{\gamma}])
-\sum_{i=1}^{n} d_i (\p\beta^i)^2
\right.
\nn\\
%& &
\left.
-({D}_0-2) (\p\ol{\gamma})^2
%-C_{\alpha\beta}(\p\Psi^\alpha)(\p\Psi^\beta)
+ e^{2\ol\gamma}\left[\sum_{i=1}^{n}e^{-2\beta^i} R_i
%-2\Lambda - 2\kappa^2\rho
\right]
\right\} ,
\ear
where $e^f$ is a dilatonic scalar field coupling to the
$D_0$-dimensional geometry on
$\ol{M}_0$.
%
%
%Analogously,
\rf{2.15} is the action of  a $\sigma$-model.
Gauging $f$ it can be
written in the form
\bear{Sf0}
{}^{(f)}\!S=\int_{\ol M_0} d^{{D}_0}x \sqrt{|\ol g^{(0)}|}
{}^{(f)}\!N^{{D}_0}
 &\! &\!
\left\{\frac{m}{2}
{}^{(f)}\!N^{-2}
\left[ R[\ol g^{(0)}] - {}^{(f)}\!G_{ij}(\p\beta^i)(\p\beta^j)
%-C_{\alpha\beta}(\p\Phi^\alpha)(\p\Phi^\beta)
\right]
\right.
\nn\\
&\! &\! \qquad \qquad
\left.
- {}^{(\E)}\!V(\beta)
\right\} \ ,
\\
\label{Nf0}
{}^{(f)}\!N\ := &\! &\! e^{\frac{f}{{D}_0-2}}\ ,
\\
\label{Vf0}
{}^{(\E)}\!V(\beta)\ := &\! &\!
m \Omega^2
\left[
%\Lambda+\kappa^2\rho
-\frac{1}{2} \sum_{i=1}^{n} R[g^{(i)}]
e^{-2\beta^i} \right]\  ,
\ear
%where the function $\Omega$ on $\ol{M}_0$ is defined as
\beq{Omega}
\Omega:=\varphi^{ \frac{1}{2-{D}_0} }\ ,
\eeq
%Note that, with $\Omega$ also the potential
where \rf{Vf0} is  gauge invariant.
The dilatonic target-space metric ${}^{(f)}\!G$,
though not even conformally flat in general,
is flat for constant $f$.
In fact, the target space is in general a conformally
homogeneous space, and in the Einstein frame a homogeneous one
(of Euclidean signature).
With
an appropriate coset representation $\rho$ of the target space,
${\frak M}={\frak G}/{\frak H}$,
the $\sigma$-model for $D_0=4$ takes the matrix form
\bear{Smatq}
{}^{(E)}\!S=\int_{\ol M_0}
 &\! &\!
\left\{\frac{m}{2}
\left[\tr \Omega\wedge *\Sigma + B \tr_{\rho}
d{\cal M}\wedge *d{\cal M}^{-1}
%-tr[C d\Phi\wedge*d\Phi]
\right]
\right.
\nn\\
&\! &\! \qquad \qquad
\left.
- {}^{(\E)}\!U({\cal M}) *1
\right\} \ ,
\ear
where $\Omega$ is the curvature $2$-form,
$\Sigma:=e\wedge e$ and $\ol g^{(0)}$ are given by
the $D_0$-dimensional soldering $1$-form $e$, and
the Hodge star is taken w.r.t. $({\ol M},g^{(0)})$.
The form \rf{Smatq} may provide also a convenient starting
point for canonical quantization, an issue not followed further here.
%
%%%%%%%%%%%%%%%%%%%%%%%%%%%%%%%%%%%%%%%%%%%%%%%%%%%%%%%%%%%%%%%%%%%%%%%%
\section{\bf The $\sigma$-model extended by scalars and $p+2$-forms}
\setcounter{equation}{0}
%%%%%%%%%%%%%%%%%%%%%%%%%%%%%%%%%%%%%%%%%%%%%%%%%%%%%%%%%%%%%%%%%%%%%%%%
We now couple the purely gravitational action \rf{2.11}
to additional matter fields of scalar and
generalized Maxwell type, i.e.
we consider now the action
\bear{2M.1}
{2\kappa^{2}}
[ S[g,\phi,F^a] -  S_{\GHY} ] =
\int_{N} d^{D}z \sqrt{|g|} \{ {R}[g]
%-2 \Lambda
- C_{\alpha\beta}g^{MN} \partial_{M} \Phi^\alpha \partial_{N} \Phi^\beta
\nn\\
- \sum_{a \in \Delta}
\frac{\eta_a}{n_a!} \exp[ 2 \lambda_{a} (\Phi) ] (F^a)^2 \}
\ear
of a self-gravitating $\sigma$ model on $M$ with
topological term $S_{\GHY}$.
%and
%optional cosmological constant $\Lambda$.
Here the $l$-dimensional target space, defined
by a vector field $\phi$ with scalar components
$\phi^\alpha$, $\alpha=1,\ldots,l$, is coupled to
several antisymmetric $n_a$-form fields $F^a$
via $1$-forms $\lambda_{a}$, $a\in\Delta$.
For consistency, we have to demand of course
that all fields are internally homogeneous.
We will see below how this gives rise
to an effective  $l+|\Delta|$-dimensional
target-space extension.
%
%With $I\subset\{1,\ldots,n\}$,
The generalized Maxwell fields
$F^a$ are located on $(n_a-1)$-dimensional world sheets
\bear{2M.20} M_{I} &:=& \prod_{i\in I} M_{i}= M_{i_1}  \times
\ldots \times M_{i_k},
\ear
of different
$(n_a-2)$-branes, labeled for each type $a$ by the sets $I$ in a
certain subset $\Omega_a\subset 2^{\{1,\ldots,n\}}$.
Variation of (\ref{2M.1}) yields the field equations \bear{2M.4}
R_{MN} - \frac{1}{2} g_{MN} R  &=&   T_{MN}  ,
\\
\label{2M.5}
C_{\alpha\beta}{\Delta}[g] \phi^\beta -
\sum_{a \in \Delta} \frac{\eta_a \lambda^{\alpha}_a}{n_a!}
e^{2 \lambda_{a}(\phi)} (F^a)^2 &=& 0 ,
\\
\label{2M.6}
\nabla_{M_1}[g] (e^{2 \lambda_{a}(\phi)}
F^{a, M_1 M_2 \ldots M_{n_a}})  &=&  0 ,
\ear
$a \in \Delta$, $\alpha=1,\ldots,l$.
In (\ref{2M.4}) the $D$-dim. energy-momentum
from \rf{2M.1} is a sum
\bear{2M.7}
T_{MN} :=  \sum_{\alpha=1}^l T_{MN}[\phi^\alpha,g]
+ \eta_a \sum_{a\in\Delta} e^{2 \lambda_{a}(\phi)} T_{MN}[F^a,g] ,
\ear
of contributions from scalar and
generalized Maxwell fields,
\bear{2M.8}
T_{MN}[\phi^\alpha,g] &:=&
C_{\alpha\beta}\p_{M} \phi^\alpha \p_{N} \phi^\alpha -
\frac{1}{2} g_{MN} \p_{P} \phi^\alpha \p^{P} \phi^\alpha ,
\\
\label{2M.9}
T_{MN}[F^a,g] &:=&
\frac{1}{n_{a}!}  \left[
- \frac{1}{2} g_{MN} (F^{a})^{2}
 + n_{a}  F^{a}_{\ M M_2 \ldots M_{n_a}} F_{\ N}^{a\ M_2 \ldots M_{n_a}}
\right] .
\ear
%We give now a sufficient criterion for the
%energy-momentum tensor \rf{2M.7} to decompose
%multidimensionally.
%
Let $W_1:= \{ i \mid i>0,\ d_i=1\}$,
% be the label set of
%$1$-dimensional factor spaces of the multidimensional
%decomposition, and set
$n_1:=| W_1 |$,
and
%. Define
\beq{Wdef}
W(a;i,j):= \{ (I,J) \mid I,J\in \Omega_{a},\
                        (I\cap J) \cup \{i\} = I \not\ni j,\
                        (I\cap J) \cup \{j\} = J \not\ni i  \} .
\eeq
Then the following criterion for MD decomposability of $T_{MN}$ holds.
\nl
{\bf Theorem:}
If for $n_1>1$ the $p$-branes satisfy
the condition
for all $a\in \Delta$, $i,j\in W_1$ with $i\neq j$,
the condition
\beq{2M.r2}
W(a;i,j) \mustbe \emptyset \quad \forall a\in \Delta \forall i,j\in W_1 ,
\eeq
then the energy-momentum \rf{2M.7}
decomposes multidimensionally without further constraints.
\hfill\mbox{$\Box$}\break
Antisymmetric fields of generalized electric type,
are given by scalar potential fields $\Phi^{a,I}$,
$a \in \Delta$, $I \in \Omega_{a}$,
which compose to a $(\sum_{a \in \Delta}|\Omega_{a}|)$-dimensional
vector field $\Phi$.
Magnetic type fields are just given as the duals of
appropriate  electric ones.
\bear{elmagF}
F^{e,I}&=& d\Phi^{e,I}\wedge\tau(I) \\
F^{m,I}&=& e^{-2\lambda_a(\phi)} * (d\Phi^{m,I}\wedge\tau(J)) .
\ear
%In the Einstein frame, the
Hence, the action \rf{2M.1}%then
reduces to
a $\sigma$-model on $M_0$ with extended
$(n + l + \sum_{a\in \Delta}| \Omega_{a}|)$-dimensional target space
and dilatonic potential (\ref{Vf0}).

For convenience, let us introduce the topological numbers
\beq{topnum}
l_{jI} := - \sum_{i \in I} D_i \delta^i_j ,
\qquad j =1,\ldots,n ,
\eeq
and with $N := n + l$ define
and define a $N \times |S|$-matrix
\beq{Lmat}
L = \left(L_{As} \right)
=
\left( \begin{array}{cc}
            &L_{i s} \\
             &L_{\alpha s}
             \end{array}
\right)
:=
\left( \begin{array}{cc}
            &l_{i I} \\
             &\lambda_{\alpha a}
             \end{array}
\right) ,
\eeq
a $N$-dimensional vector field
$(\sigma^A) := (\beta^i, \phi^{\alpha})$, $A=1,\ldots,n,n+1,\ldots,N$,
composed by dilatonic and matter scalar fields, and
a non-degenerate (block-diagonal) $N \times N$-matrix
\beq{Ghat}
\hat{G} = \left(\hat{G}_{AB} \right)    =
                             \left(
                              \begin{array}{cc}

                               G_{ij} &  0 \\
                                   0   &  C_{\alpha \beta}
                               \end{array}
                          \right) .
\eeq
With these definitions, the $\sigma$-model %\rf{Sf0M}
assumes the form
\bear{SGhat}
S_{0} =
\int_{M_0} \Bigl\{
\frac{m}{2}
\left[
{R}[g^{(0)}]
- \hat{G}_{AB} \p \sigma^A \p \sigma^B
- \sum_{s \in S} \varepsilon_s e^{2 L_{A s} \sigma^A}
(\p \Phi^s)^2
\right]
- {}^{(\E)}\!V(\sigma)
\Bigr\} \vol(g^{(0)}) \quad .
\ear

%%%%%%%%%%%%%%%%%%%%%%%%%%%%%%%%%%%%%%%%%%%%%%%%%%%%%%%%%%%%%%%%%%%%%%%%
\section{Orthobranes and the extended target space}
\setcounter{equation}{0}
%%%%%%%%%%%%%%%%%%%%%%%%%%%%%%%%%%%%%%%%%%%%%%%%%%%%%%%%%%%%%%%%%%%%%%%%
Now we present a class of solutions with
$\Ric[g^{(0)}]  = 0$, ${}^{(\E)}\!V=0$ and all fields given
in terms of harmonic functions
such that the field equations reduce to
%where the field equations read
\bear{feq1}
0  =
\hat{G}_{AB} \p_{\mu} \sigma^A \p_{\nu} \sigma^B
+ \sum_{s \in S} \eps_s e^{2 L_{As} \sigma^A}
\p_{\mu} \Phi^s  \p_{\nu} \Phi^s .
%&\qquad&
%\mu,\nu=1,\ldots, D_0 ,
%\\
%\label{feq2}
%\hat{G}_{AB} {\btu}[g^{(0)}] \sigma^B
%-  \sum_{s \in S} \eps_s L_{As} e^{2 L_{Cs} \sigma^C} (\p \Phi^s)^2 = 0 ,
%&\qquad&
%A = 1,\ldots, N ,
%\\ \label{feq3}
%\p_{\mu} \left( \sqrt{|g^{(0)}|} g^{{(0)} \mu \nu} e^{2 L_{As}
%\sigma^A} \p_{\nu} \Phi^s \right) = 0 ,
%&\qquad&
%s \in S .
\ear
For the Abelian part of the target space metric
we set $(\hat{G}^{AB}) := (\hat{G}_{AB})^{-1}$ and
$ < X,Y > := X_A \hat{G}^{AB} X_{B}$.
Let us now consider vectors
%\beq{i3.20}
$L_{s}$, $s\in S$,  % = (L_{As}) \in \R^N
%\eeq
from \rf{Lmat}   (depending only on dimensions and couplings).
$S$ is called an {\em orthobrane} index set,
iff there exists a family of real non-zero coefficients
$\{\nu_s\}_{s \in S}$, such that
\beq{i3.19}
< L_s,L_r > =
(L^{T} \hat{G}^{-1} L)_{sr} = - \eps_s (\nu_s)^{-2}\delta_{sr} ,
\qquad s,r \in S .
\eeq
For  $s \in S$  and  $A = 1, \ldots, N$, we set
\beq{i3.17}
\alpha^A_s := - \eps_s (\nu_s)^{2}  \hat{G}^{AB} L_{Bs} .
\eeq
{}\nl
\nl
\noindent
{\bf Theorem:}
Let  $ S$ be an orthobrane index set
with coefficients \rf{i3.17}.
If for any  $s\in S$
there is a harmonic function  $H_s > 0$ on $M_0$
then the fields
%the field configuration
\bear{i3.14}
%R_{\mu \nu}[g^{(0)}]  = 0 ,
%&\qquad&
%\mu,\nu=1,\ldots, D_0 ,
%\\
%\label
\sigma^A := \sum_{s \in S} \alpha^A_s \ln H_s ,
&\qquad&
A = 1, \ldots, N ,
\\
\label{i3.15} \Phi^s  := \frac{\nu_s}{H_s} , &\qquad& s \in S \ear
satisfy \rf{feq1} and hence the field equations.
% (\ref{feq1})-(\ref{feq3}).
\hfill\mbox{$\Box$}\break
%\nl
With ${\rm Ric}[g^{(i)}] =0$, $i=0,\ldots n$,
the general solution geometry then
%of \rf{i3.13} - \rf{i3.15}
reads
\bear{xxxxx} \nn
g&=& \biggl(
\prod_{s \in S}
H_s^{2 \alpha^0_s}
\biggr)
g^{(0)}
+ \sum_{i=1}^{n}
\biggl(
\prod_{s \in S}
H_s^{2 \alpha^i_s}
\biggr)
g^{(i)}
\\
\label{i4.1}
&=&
\left(
\prod_{(a,I)\in S}
H_{a,I}^{ \eps(I) 2 D(I) \nu^2_{a,I} }
\right)^{1/(2-D)}
%\times
\left\{
g^{(0)}
+  \sum_{i=1}^{n}
\left(
\prod_{(a,I) \in S, I \ni i}
H_{a,I}^{\eps(I) 2 \nu^2_{a,I} }
\right)
g^{(i)}
\right\} ,
\ear
\bear{i4.p}
\phi^\beta  =  \sum_{s \in S}
\alpha^\beta_{s} \ln H_s
= - \sum_{(a,I) \in S}
\eps(I) C^{\beta\gamma}\lambda_{\gamma a}
\nu^2_{a,I} \ln H_{a,I} ,
&\qquad& \beta=1,\ldots,l ,
\\
\label{i4.a1}
A^{a} = \sum_{I \in \Omega_{a}}
\frac{\nu_{a,I}}{H_{a,I}} \tau_{I} ,
&\qquad& a \in \Delta ,
\ear
%where forms $\tau_I$ are defined in \rf{2.8}.
By \rf{i3.19} parameters $\nu_s \neq 0$ and ${\lambda}_a$
%satisfy the orthobrane condition (\ref{i3.28}),
%
%Finally recall that these solutions
are subject to the  {\em orthobrane} condition
\beq{i4.2}
D(I \cap J) + \frac{D(I) D(J)}{2-D}
+ C^{\alpha\beta} {\lambda}_{\alpha a}  {\lambda}_{\beta b}
= - \eps(I) (\nu_{a,I})^{-2} \delta_{ab}\delta_{I,J}\ ,
\quad 0\neq\nu_{a,I}\in\R\ ,
\eeq
for $a, b \in \Delta$, $I \in \Omega_{a}$, $I \in \Omega_{b}$.
The latter implies
specific intersection rules for the $p$-branes.
(Historically it was by heuristic settings of such rules that
intersecting $p$-brane solutions were found.)
%Some concrete examples
%
For positive definite $(C_{\alpha\beta})$
(or $(C^{\alpha \beta})$) and $D_0 \geq 2$, \rf{i4.2} implies
\beq{i3.43}
\eps(I)=-1 ,
\eeq
for all $I \in \Omega_{a}$, $a \in \Delta$.
Then, the restriction $g_{\vert M_{I}}$ of the metric \rf{i4.1}
to a membrane manifold $M_{I}$ has an odd number
of negative eigenvalues,
i.e. linearly independent time-like directions.
However, if the metric $(C_{\alpha\beta})$ in the space of scalar fields
is not positive definite,
then \rf{i3.43} may be violated for sufficiently negative
$C^{\alpha\beta} {\lambda}_{\alpha a}  {\lambda}_{\beta b}<0$.
%%%%%%%%%%%%%%%%%%%%%%%%%%%%%%%%%%%%%%%%%%%%%%%%%%%%%%%%%%%%%%%%%%%%%%%%%
%\section{\bf Target space structure}
%\setcounter{equation}{0}
%%%%%%%%%%%%%%%%%%%%%%%%%%%%%%%%%%%%%%%%%%%%%%%%%%%%%%%%%%%%%%%%%%%%%%%%
%
%{\bf Theorem:}

The extended target space $({\frak M},{\frak g})$ is a homogeneous
space, and the following holds.
\nl
{\bf Theorem:}
The target space $({\frak M},{\frak g})$ is locally symmetric
if and only if $<L^s,L^r>_{\hat G}(L^s-L^r)=0$.

Proof: ${\frak Riem}$, the Riemann tensor of
$({\frak M},{\frak g})$, is locally symmetric, if and only if
\beq{locsym}
\nabla{\frak Riem}=0 ,
\eeq
where $\nabla$
denotes the covariant derivative w.r.t ${\frak g}$.
The only non-trivial equations \rf{locsym} are
\beq{DRiem}
\nabla_p {\frak R}_{s r q A} =
k_{psrq}
<L^{s}, L^{r}>_{\hat G}(L^{r}_A - L^{s}_A)
= 0,
\quad
A=1,\ldots,N,
\quad
p,q,r,s\in S
\eeq
with
$k_{psrq}:=
\eps_{s} \eps_{r}
e^{2U^{s} + 2U^{r}}
(\delta_{p s} \delta_{r q} + \delta_{p r} \delta_{s q})$
nonzero for fixed $s,r$.
\hfill\mbox{$\Box$}\break
Recall that with $L$ also the target space structure is
given by dimensions and couplings.
%%%%%%%%%%%%%%%%%%%%%%%%%%%%%%%%%%%%%%%%%%%%%%%%%%%%%%%%%%%%%%%%%%%%%%%%
\section{\bf Static, spherically symmetric solutions}
\setcounter{equation}{0}
%%%%%%%%%%%%%%%%%%%%%%%%%%%%%%%%%%%%%%%%%%%%%%%%%%%%%%%%%%%%%%%%%%%%%%%%
let us now examine static, spherically symmetric, multidimensional
space-times with
\beq{8}
        M=M_{-1} \times M_0\times M_1\times \cdots \times M_N,
        \cm \dim M_i=d_i, \quad i=0,\ldots, N ,
\eeq
where $M_{-1}\subset {\R}$ corresponds to a radial coordinate $u$,
$M_0 = S^2$ is a 2-sphere, $M_1\subset\R$ is time, and $M_i,\ i>1$
are internal factor spaces.
The metric is assumed correspondingly to be
\beq{9}
     ds^2 =
            \e^{2\alpha(u)}du^2 + \sum_{i=0}^N\e^{2\beta_i(u)}ds_i^2
%\nn\\
%            \al\equiv\al
=                -\e^{2\gamma(u)}dt^2+\e^{2\alpha(u)}du^2
              +\e^{2\beta_0(u)}d\Omega^2
              +\sum_{i=2}^N\e^{2\beta_i(u)}ds_i^2  ,
\eeq
where
$ds_0^2 \equiv d\Omega^2=d\theta+\sin^2\theta\, d\phi^2$
is the line element on $S^2$,
$ds_1^2 \equiv -dt^2$ with $\beta_1 =: \gamma$,
and $ds_i^2$, $i> 1$, are $u$-independent line elements of internal
Ricci-flat spaces of arbitrary dimensions $d_i$ and signatures
$\eps_i$.

For simplicity in the following we consider only one single
scalar field denoted as $\varphi$.
An electric-type $p+2$-form $F_{\e I}$ has a domain
given by a product manifold $M_I = M_{i_1} \times \cdots \times M_{i_k}$.
%\beq{MI}
%\eeq
A magnetic-type $F$-form of arbitrary rank $k$ may be defined
as a form on a  domain $M_{\oI}$ with $\oI \eqdef I_0 - I$,
dual to an electric-type form,
\beq{Fm}
     F_{\m I,\,M_1\ldots M_k}
        = \e^{-2\lambda\varphi} (*F)_{\e I,\,M_1\ldots M_k}
        \equiv \e^{-2\lambda\varphi}
        \frac{\sqrt{g}}{k!} \eps_{M_1\ldots M_k N_1\ldots N_{D-k}}
        F_{\e I}^{N_1\ldots N_{D-k}},
\eeq
where $*$ is the Hodge operator and
$\eps$ is the totally antisymmetric Levi-Civita symbol.

For simplicity we now considering a just a single $n$-form,
i.e. a single electric type and a single dual magnetic component,
whence $k=n$ in (\ref{Fm}) and
\beq{12}
        d(I) = n-1   \quad \mbox{for} \quad F_{\e I}, \cm
        d(I) = d(I_0)-n = D-n-1 \quad \mbox{for} \quad F_{\m I}.
\eeq

All fields must be compatible
with spherical symmetry and staticity.
Correspondingly, the vector $\varphi$ of scalars and
the $p+2$-forms valued fields depend (besides on their domain as forms)
on the radial variable $u$ only.
Furthermore,
the domain of the electric form $F_{\e I}$
does not include the sphere $M_0 = S^2$, and
%$F_{\e I}$
%is specified by a
%$u$-dependent potential form,
\beq{Ue}
        F_{\e I,\,u L_2\ldots L_n} = \bdy_{\,[u}U_{L_2\ldots L_n ]} \cm
        U = U_{L_2,\ldots,L_{n}} dx^{L_2}\wedge\ldots\wedge dx^{L_n} .
\eeq

Since the time manifold $M_1$ is a factor space of $M_I$, the form
(\ref{Ue}) describes an electric $(n-2)$-brane in the remaining
subspace of $M_I$. Similarly \rf{Fm} describes a magnetic
$(D-n-2)$-brane in $M_I$.

Let us label all nontrivial components of $F$ by a collective
index $s = (I_s,\chi_s)$, where $I=I_s\subset I_0$ characterizes the
subspace of $M$ as described above and
$\chi_s:=\pm 1$ for the electric resp. magnetic case.
Let us assume Ricci-flat internal spaces.
With spherical symmetry and staticity
all fields become independent of $M_{0}$ and $M_{1}$,
and the variation reduces further from $\ol M_0$ to the
radial manifold $M_{-1}$ only.
The generalized Maxwell equations %(\ref{4})
give
%(with (\ref{3.1})):
\bear{3.2}
        F_{\e I}^{uM_2\ldots M_n}
                \eql Q_{\e I}\e^{-2\alpha - 2\lambda\varphi},
                        \qquad \ \ Q_{\e I}= \const,
\\\label{3.3}
        F_{\m I,\, uM_1\ldots M_{d(\oI)}}
                \eql Q_{\m I} \sqrt{|g_{\oI}|},\qquad
                        \qquad  Q_{\m I}= \const,
\ear
where $|g_{\oI}|$ is the determinant of the $u$-independent
part of the metric of $M_{\oI}$ and $Q_s$ are charges.
These solutions provide then the energy momentum tensors,
of the electric and magnetic $p+2$-forms written in matrix form,
\bearr{3.4x}
    \e^{2\alpha}(T_M^N [F_{\e I}])
          = -\half \eta_F \eps(I) \Qie^2 \e^{2y_{\e I}}
                 \diag\bigl(+1,\ \pO_I,\ \mO_{\oI} \bigr);
\nnnv
    \e^{2\alpha}(T_M^N [F_{\m I}])
          = \half \eta_F \eps(\oI) \Qim^2
              \e^{2y_{\m I}} \diag\bigl(1,\ \pO_I,\ \mO_{\oI} \bigr),
\ear
where the first position belongs to $u$
and $f$ operating over $M_J$ is denoted by
$[f]\mathstrut_J$.
The functions $y_s (u)$ are
\beq{3.5}
        y_s (u) = \sigma (I_s) - \chi_s \lambda\varphi.
\eeq
The functions $y_s(u)$ (\ref{3.5}) can be represented as scalar
products in $V$ (recall that $s = (I_s, \chi_s)$):
\bear{3.16}
        y_s (u)   = Y_{s,A}  x^A,    \inch
        (Y_{s,A}) = (d_i\delta_{iI_s}, \ \  -\chi_s \lambda),
\ear
where $\delta_{iI} := \sum_{j\in I}\delta_{ij}$
is an indicator for $i$ belonging to $I$ (1 if $i\in I$ and 0 otherwise).

The contravariant components of $Y_s$ are
%found using the matrix
%$\oG^{AB}$ inverse to $\oG_{AB}$:
\bearr{3.18}
%                                                                                                 \label{3.19}
        (Y_s{}^A) =
        \biggl(\delta_{iI_s}-\frac{d(I_s)}{D-2}, \ \ -\chi_s \lambda\biggr),
\ear
and the scalar products of different $Y_s$
%,
%whose values are of primary
%importance for the integrability of our system,
are
\beq{3.20}
        Y_{s,A}Y_{s'}{}^A = d(I_s \cap I_{s'})
                                        - \frac{d(I_s)d(I_{s'})}{D-2}
                        + \chi_s\chi_{s'} \lambda^2.
\eeq
%
%
%
%
%%%%%%%%%%%%%%%%%%%%%%%%%%%%%%%%%%%%%%%%%%%%%%%%%%%%%%%%%%%%%%%%%%%%%
%\section{Purely EM black hole solutions}   % S.6
%\setcounter{equation}{0}
%%%%%%%%%%%%%%%%%%%%%%%%%%%%%%%%%%%%%%%%%%%%%%%%%%%%%%%%%%%%%%%%%%%%%%%%
It was shown\cite{BKR} that quasiscalar components of the
$F$-fields are incompatible with orthobrane \bhs. Therefore we here
consider only two $F$-field components,
%Type \E\ and Type \M\
%according to the classification above. They will be electric as
$F_\e$ and $F_\m$ with
%denote the
corresponding sets
%$I_s\subset I_0$ as
$I_\e$ and $I_\m$. Then a minimal toy configuration (\ref{8}) of the
manifold $M$
%compatible with an arbitrary choice of $I_s$
has the
%following
form
\beq{6.1}
        N=5,\cm I_0 = \{0,1,2,3,4,5\},  \cm
        I_e = \{1,2,3\},        \cm
        I_m = \{1,2,4\},
\eeq
so that
\bearr
        d(I_0) = D-1, \qquad d(I_\e) = n-1, \qquad d(I_\m)= D-n-1,  \qquad
        d(I_\e \cap I_\m) = 1 + d_2;
\nnnv
\label{6.2}
        d_1=1, \inch   d_2+d_3 = d_3 + d_5 = n-2.
\ear
The relations (\ref{6.2}) show that, given $D$ and $d_2$, all $d_i$ are
known.
The configuration  corresponds to an electric
($n-2$)-brane located on the subspace $M_2\times M_3$ and a
magnetic ($D-n-2$)-brane on the subspace $M_2\times M_4$.
Their intersection dimension $d_{\ints}=d_2$ turns out to
determine qualitative  properties of the solutions.
%
%%%%%%%%%%%%%%%%%%%%%%%%%%%%%%%%%%%%%%%%%%%%%%%%%%%%%%%%%%%%%%%%%%%%%%%%
\section{Orthobrane black hole solutions}
\setcounter{equation}{0}
%%%%%%%%%%%%%%%%%%%%%%%%%%%%%%%%%%%%%%%%%%%%%%%%%%%%%%%%%%%%%%%%%%%%%%%%
Let us consider now orthobranes, where vectors $Y_s$ are mutually orthogonal with
        respect to the metric $\oG_{AB}$, i.e.
\beq{4.1}
        Y_{s,A}Y_{s'}{}^A = \delta_{ss'}N_s^2 .
\eeq
%
%%%%%%%%%%%%%%%%%%%%%%%%%%%%%%%%%%%%%%%%%%%%%%%%%%%%%%%%%%%%%%%%%%%%%%%%
%%%%%%%%%%%%%%%%%%%%%%%%%%%%%%%%%%%%%%%%%%%%%%%%%%%%%%%%%%%%%%%%%%%%%%%%
For the minimal configuration  \rf{6.1}
%-\rf{6.3},
the orthogonality condition (\ref{4.1})  reads
\beq{6.4}
        \lambda^2=  d_2+1 -\frac{1}{D-2}(n-1)(D-n-1) .
\eeq
%In particular, in
For dilaton gravity $n=2,\ d_2=0$, and the
%integrability condition
\rf{6.4} reads $\lambda^2 = 1/(D-2)$.
Some examples of
configurations satisfying the orthogonality condition (\ref{6.4})
in the purely topological case $\lambda=0$ are summarized in Table
\ref{tab2} (including the values of the constants $B$ and $C$ from
(\ref{6.12}) ).
%In this case (\ref{6.4}) is just a Diophantus
%equation for $D$, $n$ and $d_2$.
%
% Tabelle als Gleitumgebung
\begin{table}[htb!]
\caption{Orthobrane solutions with $\lambda=0$}
\label{tab2}
\begin{center}
\begin{tabular}{|l|l|l|l|l|l|l|}
\hline
                &&&&&& \\
     & $n$   &  $d(I_{\e})$  &  $d(I_{\m})$  & $d_2$ & $B$ & $C$ \\
                        &&&&&& \\
      \hline
      &&&&&& \\
 $D = 4m + 2$        & 2m{+}1  & 2m &  2m &\ \ m{-}1 & 1/m  & 1/m\\
 ($m\in \N$) &&&&&& \\
                        &&&&&&                             \\
 \qquad D= 11          &  4    &  3 &   6  &    1 & 2/3  &  1/3   \\
                             &  7    &  6 &   3  &    1 & 1/3  &  2/3   \\
                        &&&&&&                             \\
\hline
\end{tabular}
\end{center}
\end{table}
%
%
%
%Let us now consider the case where
%(\ref{5.4}) and (\ref{5.5}) with $\delta_{1I_s} =1$ hold.
After a transformation
$u\mapsto R$, to isotropic coordinates given by the relation
\beq {6.7}
     \e^{-2ku}=1-2k/R ,
\eeq
we obtain the solution
\bearr{6.8}
     ds^2 = -\frac{1-2k/R}{P_\e^{B}P_\m^{C}} dt^2
                +P_\e^{C}P_m^{B}
     \left(\frac{dR^2}{1-2k/R}+R^2 d\Omega^2\right)
                                +\sum_{i=2}^5 \e^{2\beta_i(u)}ds_i^2 ,
\yyy\label{6.9}
     \e^{2\beta_2} = {P_\e}^{-B}{P_\m}^{-C},
\cm
     \e^{2\beta_3} = \left({P_\m}/{P_\e}\right)^{B},
\nnnv
     \e^{2\beta_4} = \left({P_\e}/{P_\m}\right)^{C} ,
\cm
     \e^{2\beta_5} = {P_\e}^{C}{P_\m}^{B} ,
\yyy
\label{6.10}
 e^{2\lambda\varphi} =
     ({P_\e}/{P_\m})^{2\lambda^2/(1+d_2)} ,
 \yyy \label{6.11}
     F_{01M_3\ldots M_n}=-{Q_\e}/{(R^2 P_\e)} ,
\cm
     F_{23M_3 \ldots M_n}=Q_\m \sin\theta,
\ear
with the notations
\bearr{6.12}
        P_{\e,\m} = 1+ p_{\e,\m}/R, \qquad
        p_{\e,\m} = \sqrt{k^2+ (1+d_2) Q_{\e,\m}^2} -k;
\nnnv
        B = \frac{2(D-n-1)}{(D-2)(1+d_2)},
\cm
        C = \frac{2(n-1)}{(D-2)(1+d_2)}.
\ear
For a  static, spherical BH
one can define a Hawking temperature $T_H:= \kappa/2\pi$
as given by the surface gravity $\kappa$.
With a generalized Komar integral
%(see e.g. \cite{Heusler})
\beq{Komar}
M(r):=-\frac{1}{8\pi}
\int_{S_{r}} * d\xi
\eeq
over the time-like Killing form $\xi$,
the surface gravity can be  evaluated as
\beq{6.15}
     \kappa   = M(r_H)/(r_H)^2
     =(\sqrt{|g_{00}|})'\Big/\sqrt{g_{11}}\biggr|_{r=r_H}
     =\e^{\gamma-\alpha}|\gamma'|\, \biggr|_{r=r_H}\,.
\eeq
Substituting $g_{00}$ and $g_{11}$ from \rf{6.8}, one obtains
\beq{6.16}
     T_H = \frac{1}{2\pi k_{\rm B}} \frac{1}{4k}
         \left[\frac{4k^2}{(2k+p_\e)(2k+p_\m)}\right]^{1/(d_2+1)},
\eeq
with extreme limit $k\to 0$ as follows:
If $d_2=0$ and both charges are nonzero, $T_H\to 0$,
if $d_2=1$ and both
charges are nonzero, $T_H\to \frac{1}{4\pi k_{\rm B}\sqrt{ p_\e p_\m}}$,
and $T_H\to \infty$ otherwise.
Remarkably, $T_H$ critically depends on $d_2+1$,
which is the dimension of the intersection of the
$p$-brane world sheets.
%rather than the whole space-time
%dimension $D$.
%%%%%%%%%%%%%%%%%%%%%%%%%%%%%%%%%%%%%%%%%%%%%%%%%%%%%%%%%%%%%%%%%%%%%%%%%%%%%%
\section{Black hole solution for $Q_\e^2 = Q_\m^2$}  %
\setcounter{equation}{0}
%%%%%%%%%%%%%%%%%%%%%%%%%%%%%%%%%%%%%%%%%%%%%%%%%%%%%%%%%%%%%%%%%%%%%%%%%%%%%
%%%%%%%%%%%%%%%%%%%%%%%%%%%%%%%%%%%%%%%%%%%%%%%%%%%%%%%%%%%%%%%%%%%%%%%%
In this degenerate case, solutions can be found which need not satisfy
the orthobrane condition (\ref{4.1}).
Let us suppose that two functions (\ref{3.5}), say, $y_1$ and $y_2$,
coincide
up to an addition of a constant
(which may be then absorbed by re-defining a
charge $Q_1$ or $Q_2$) while
corresponding vectors $Y_1$ and $Y_2$
are neither coinciding, nor orthogonal (otherwise we would have
the previously considered situation). Substituting $y_1\equiv y_2$ into
(\ref{3.16}), one obtains
\beq{4.10}
        (Y_{1,A} - Y_{2,A})x^A =0 .
\eeq
This is a constraint reducing the number of
independent unknowns $x^A$.
%\eeq
The event horizon occurs at $u=\infty$. After the transformation
(\ref{6.7}) the metric takes the form
\bear{6.23}
     ds^2_D \eql
            - \frac{1-2k/R}{(1+p/R)^{2\nu}}dt^2
     +(1+p/R)^{2\nu}
                  \left(\frac{dR^2}{1-2k/R}+R^2d\Omega^2\right)
\nnnv
\inch\cm
+(1+p/R)^{-2\nu} ds_2^2 +ds_3^2 + ds_4^2 + (1+p/R)^{2\nu}ds_5^2
\ear
with the notation
\beq{6.24}
      p = \sqrt{k^2 + (1+d_2)Q^2}-k.
\eeq
The fields $\varphi$ and $F$ are determined by the relations
\beq{6.25}
     \varphi\equiv 0\ ,
\cm
     F_{01L_3 \ldots L_n} = -\frac{Q} {R^2(1+p/R)},
\cm
     F_{23L_3 \ldots L_n} = Q \sin \theta.
\eeq
%
%\beq{6.26}
%G_N M = k +p/(1+d_2),
%\eeq
The Hawking temperature can be calculated as before,
\beq{6.27}
     T_H = \frac{1}{2\pi k_{\rm B}}
             \frac{1}{4k}\left(\frac{2k}{2k+p}\right)^{2/(d_2+1)}.
\eeq
Here  $d_2=0$ is the Reissner-Nordstr\"om case
with $T_H\to 0$ for $k\to 0$. In the latter limit, for $d_2=1$,
$T_H\to \frac{1}{4\pi k_{\rm B}{ p}}$,
and $T_H\to \infty$ otherwise.
Again $T_H$ critically depends
%on the space-time dimension
%$D$, but depends on the $p$-brane intersection dimension
on $d_2$.

\end{document}